# Modeling the relationship between regional activation and functional connectivity during wakefulness and sleep

Ignacio Perez Ipiña[1], Patricio Donnelly Kehoe[2,3,4,5], Morten Kringelbach[6,7], Helmut Laufs[8], Agustín Ibañez[2,9,10,11,12], Gustavo Deco[13,14], Yonatan Sanz Perl[1,2,9*], Enzo Tagliazucchi[1,2*]

[1]Department of Physics, University of Buenos Aires, Argentina

[2]National Scientific and Technical Research Council (CONICET), Buenos Aires, Argentina

[3]Centro Internacional Franco Argentino de Ciencias de la Información y de Sistemas (CIFASIS), National Scientific and Technical Research Council (CONICET), Rosario, Argentina

[4]Laboratory for System Dynamics and Signal Processing, Universidad Nacional de Rosario, Argentina

[5]Laboratory of Neuroimaging and Neuroscience (LANEN), INECO Foundation Rosario, Rosario, Argentina

[6]Department of Psychiatry, University of Oxford, Oxford, UK

[7]Center for Music in the Brain (MIB), Dept. of Clinical Medicine, Aarhus University, Denmark

[8]Department of Neurology, University of Kiel, Kiel, Germany

[9]Laboratory of Experimental Psychology and Neuroscience (LPEN), Institute of Cognitive and Translational Neuroscience (INCYT), INECO Foundation, Favaloro University, Buenos Aires, Argentina

[10]Centre of Excellence in Cognition and its Disorders, Australian Research Council (ARC), Sydney, Australia

[11]Center for Social and Cognitive Neuroscience (CSCN), School of Psychology, Universidad Adolfo Ibáñez, Santiago, Chile

[12]Universidad Autónoma del Caribe, Barranquilla, Colombia

[13]Center for Brain and Cognition, Computational Neuroscience Group, Department of Information and Communication Technologies, Universitat Pompeu Fabra, Barcelona, Spain

[14]Institució Catalana de la Recerca i Estudis Avançats (ICREA), Universitat Pompeu Fabra, Barcelona, Spain

*These authors contributed equally to this work

**Abstract**

Global brain activity self-organizes into discrete patterns characterized by distinct behavioral observables and modes of information processing. The human thalamocortical system is a densely connected network where local neural activation reciprocally influences coordinated collective dynamics. We introduce a semi-empirical model to investigate the relationship between regional activation and long-range functional connectivity in the different brain states visited during the natural wake-sleep cycle. Our model combines functional magnetic resonance imaging (fMRI) data, *in vivo* estimates of structural connectivity, and anatomically-informed priors that constrain the independent variation of regional activation. As expected, priors based on functionally coherent networks resulted in the best fit between empirical and simulated brain activity. We show that sleep progressively divided the cortex into regions presenting opposite dynamical behavior: frontoparietal regions approached a bifurcation towards local oscillatory dynamics, while sensorimotor regions presented stable dynamics governed by noise. In agreement with human electrophysiological experiments, sleep onset induced subcortical deactivation and uncoupling, which was subsequently reversed for deeper stages. Finally, we introduced external forcing of variable intensity to simulate external perturbations, and identifiedthe key regionsespecially relevant for the recovery of wakefulness from deep sleep. Our model represents sleep as a state where long-range decoupling and regional deactivation coexist with the latent capacity for a rapid transition towards wakefulness. The mechanistic insights provided by our simulations allow the *in silico* parametric exploration of such transitions in terms of external perturbations, with potential applications for the control of physiological and pathological brain states.

**Introduction**

The human brain is a complex system comprised by $10^{10}$ nonlinear units (neurons) interacting in $10^{15}$ sites (synapses) (Sporns et al., 2005). Considering such an astonishing level of complexity and heterogeneity, it is surprising that the global dynamics of the brain self-organize into a discrete set of well-defined states (Tart, 1971; Tassi and Muzet, 2001). Prime examples are the physiological transitions that take place during the wake-sleep cycle, as well as those elicited by pharmacological manipulations or as the consequence of brain injury. In spite of constituting some of the most commonplace experiences available to human beings, very little is known about the collective properties of the brain that allow rapid transitions between stable patterns with distinct modes of information processing (Stevner et al., 2019). The development of a theory of brain states as emergent properties of local neural dynamics could have far-reaching consequences both in basic and clinical neuroscience, since some of the most devastating neurological conditions result from pathological stabilization into dysfunctional states, or destabilization of physiological states.

The use of non-invasive neuroimaging tools, such as functional magnetic resonance imaging (fMRI), has contributed enormously to the characterization of brain activity during different brain states (Boly et al., 2008). It is now widely accepted that the baseline of the brain is dynamic and presents non-trivial spatiotemporal correlations (functional connectivity [FC]) within well-defined functional systems, known as resting state networks (RSNs) (Damoiseaux et al., 2006). Ample evidence supports the neurobiological relevance of RSNs. Multimodal imaging studies revealed significant correlations between metabolic activity at different RSNs and electrophysiological activity simultaneously measured with electroencephalography (EEG) (Laufs et al., 2003; Mantini et al., 2007). The configuration of different RSNs reflects the level of conscious awareness (Heine et al,.2012; Demertzi et al., 2015), ongoing cognitive processes (Waites et al., 2005; Yan et al., 2009; Cole et al., 2016), and is modified by a diversity of neuropsychiatric conditions (Broyd et al., 2009).Taken together, this evidence strongly suggests that the components of RSNs play a key role in the self-organization of brain activity into different states, yet our understanding of the relationship between RSNs and the emergence ofcoordinated collective dynamics is limited by the purely correlational nature of neuroimaging data.

While RSNs are defined in functional terms, they are also constrained by the underlying structural connectivity (SC) of the brain (Greicius et al., 2009; Van Den Heuvel et al., 2009; Haimovici et al., 2013), which contributes towards their partial preservation during states of reduced conscious awareness, including deep sleep (Tagliazucchi et al., 2016), disorders of consciousness (Demertzi et al., 2019), and the acute effects of general anesthetics (Barttfeld et al., 2015). Thus, the collective dynamics captured by FC arise from the interplay between local dynamics and coupling by long-range SC. To disentangle these different contributions, recent studies have introduced whole-brain models of low dimensionality combining fMRI data with SC estimated using diffusion tensor imaging (DTI) (Deco et al., 2011; Deco et al., 2013; Cabral et al., 2014, Cabral et al., 2017). Applied to the different stages of human sleep as paradigmatic cases of discrete brain states, these models showed promise to evaluate hypotheses about the nature of spontaneous brain activity across the wake-sleep cycle (Jobst et al., 2017), as well as to predict its behavior under different simulated stimulation protocols (Deco et al., 2018a).

While low dimensionality models are useful to provide conceptual insights, they are limited to investigate regional dynamics that can be traced back to known neurophysiological systems embedded within different RSNs. In the present work, we developed whole-brain models of fMRI activity recorded during the human wake-sleep cycle incorporating regional parameters constrained by anatomical priors. Modeling the dynamics of each region as a supercritical Hopf bifurcation, these parameters govern the transition from an activated synchronous state towards a quiescent stable state with noisy fluctuations. We performedparameter fitting to infer how simultaneous contributions of cortical and subcortical functional networks were combined to achieve and maintain differentiated brain states. We expected that fitting our model to fMRI FC revealedassociations between specific sleep stages and changes in the parameters that determine the qualitative behavior of RSNs, which cannot be inferred from fMRI recordings alone.  Beyond our specific application to the human wake-sleep cycle, our model links different sources of empirical data to narrow the space ofpossible mechanistic explanations of changes in brain activity arising during different states of consciousness, and allows the rehearsal of *in silico* perturbations that would be challenging or even impossible to implement experimentally*in vivo*.

**Materials and Methods**

*Participants and experimental protocol*

A large cohort of 63 healthy subjects participated in the original experiments (36 females, mean ± SD age of 23.4 ± 3.3 years). Written informed consent was obtained from all subjects. The experimental protocol was approved by the local ethics committee (Goethe-Universität Frankfurt, Germany, protocol number: 305/07). The subjects were reimbursed for their participation. All experiments were conducted in accordance with the relevant guidelines and regulations, and the Declaration of Helsinki.

The participants entered the scanner in the evening and underwent a resting state fMRI session with simultaneous EEG acquisition lasting for 52 minutes. Participants were not instructed to fall asleep, but were asked to relax, close their eyes and not actively fight the onset of sleep. Lights were dimmed in the scanner room and subjects were shielded from scanner noise using earplugs. The day of the study all participants reported a wake-up time between 5:00 AM and 11:00 AM, and a sleep onset time between 10:00 PM and 2:00 AM for the night prior to the experiment. Sleep diaries confirmed that these values were representative of the 6 days prior to the experiment.

Sleep staging was based on simultaneously acquired polysomnography data and performedaccording to the standard rules of the American Academy of Sleep Medicine (Iber et al., 2007). For the included participants the mean (±SD) durations of contiguous sleep epochs were 12.37 ± 6.61 minutes for wakefulness, 8.52 ± 2.83 minutes for N1, 14.69 ± 5.72 minutes for N2 and 16.56 ± 8.39 minutes for N3.

*Simultaneous fMRI and EEG data collection*

EEG via a cap (modified BrainCapMR, Easycap, Herrsching, Germany) was recorded continuously during fMRI acquisition (1505 volumes of T2*-weighted echo planar images, TR/TE = 2080 ms/30 ms, matrix 64 × 64, voxel size 3 × 2 × 2 mm3, distance factor 50%; FOV 192 mm$^2$) with a 3 T Siemens Trio (Erlangen, Germany). An optimized polysomnographic setting was employed (chin and tibial EMG, ECG, EOG recorded bipolarly [sampling rate 5 kHz, low pass filter 1 kHz] with 30 EEG channels recorded with FCz as the reference [sampling rate 5 kHz, low pass filter 250 Hz]. Pulse oxymetry and respiration were recorded via sensors from the Trio [sampling rate 50 Hz]) and MR scanner compatible devices (BrainAmp MR+, BrainAmpExG; Brain Products, Gilching, Germany), facilitating sleep scoring during fMRI acquisition.

MRI and pulse artefact correction were performed based on the average artefact subtraction (AAS) method (Allen et al., 1998) as implemented in Vision Analyzer2 (Brain Products, Germany) followed by objective (CBC parameters, Vision Analyzer) ICA-based rejection of residual artefact-laden components after AAS resulting in EEG with a sampling rate of 250 Hz. EEG artefacts due to motion were detected and eliminated using an ICA procedure implemented in Vision Analyzer2. Sleep stages were scored manually by an expert according to the AASM criteria (Iber et al., 2007). Previous publications based on this dataset can be consulted for further details (e.g. Tagliazucchi et al., 2012).

*fMRI data preprocessing*

Using Statistical Parametric Mapping (SPM8, www.fil.ion.ucl.ac.uk/spm),Echo Planar Imaging (EPI) data were realigned, normalised (MNI space) and spatially smoothed (Gaussian kernel, 8 mm$^3$ full width at half maximum). Data was re-sampled to 4 × 4 × 4 mm resolution to facilitate removal of noise and motion regressors. Note that re-sampling introduces local averaging of Blood Oxygen Level Dependent (BOLD) signals, which wereeventually averaged over larger cortical and sub-cortical regions of interest determined by the automatic anatomic labelling (AAL) atlas (Tzourio-Mazoyer et al., 2002). Cardiac, respiratory (both estimated using the RETROICOR method [Glover et al., 2000]) and motion-induced noise (three rigid body rotations and translations, as well as their first 3 temporal derivatives, resulting in 24 motion regressors) (Friston et al., 1996) were regressed out by retaining the residuals of the best linear least square fit. Data was band-pass filtered in the range 0.01–0.1 Hz (Cordes et al.,2001) using a sixth order Butterworth filter.

*DWI data collection and processing*

The structural connectome was obtained applying diffusion tensor imaging (DTI) to diffusion weighted imaging (DWI) recording from 16 healthy right-handed participants (11 men and 5 women, mean age: 24.75 ± 2.54), recruited through the online recruitment system at Aarhus University. Subjects with psychiatric or neurological disorders (or a history thereof) were excluded from participation. The MRI data (structural MRI, DTI) were recorded in a single session on a 3 T Siemens Skyra scanner at CFIN, Aarhus University, Denmark. The following parameters have been applied for the structural MRI T1 scan: voxel size of 1 mm$^3$; reconstructed matrix size 256 × 256; echo time (TE) of 3.8 ms and repetition time (TR) of 2300 ms.

The DWI data were collected using the following parameters: TR = 9000 ms, TE = 84 ms, flip angle = 90°, reconstructed matrix size of 106 × 106, voxel size of 1.98 × 1.98 mm with slice thickness of 2 mm and a bandwidth of 1745 Hz/Px. Furthermore, the data were recorded with 62 optimal nonlinear diffusion gradient directions at b = 1500 s/mm$^2$. Approximately one non-diffusion weighted image (b = 0) per 10 diffusion-weighted images was acquired. Additionally, the DTI images were recorded with different phase encoding directions. One set was collected applying anterior to posterior phase encoding direction and the second one was acquired in the opposite direction. The AAL template was used to parcellate the entire brain into 90 regions (76 cortical regions and 14 subcortical regions). The parcellation contained 45 regions in each hemisphere. To co-register the EPI image to the T1-weighted structural image, the linear registration tool from the FSL toolbox (www.fmrib.ox.ac.uk/fsl, FMRIB, Oxford)(Jenkinson et al., 2002) was employed. The T1-weighted images were co-registered to the T1 template of ICBM152 in MNI space. The resulting transformations were concatenated, inversed and further applied to warp the AAL template from MNI space to the EPI native space, where the discrete labelling values were preserved by applying nearest-neighbor interpolation. Brain parcellations were conducted in each individual's native space. The acquired DTI data was used to generate the SC maps for each participant. The two recorded datasets were processed, each with different phase encoding to optimize signal in problematic regions. The SC maps were constructed following a three-step process. First, the regions of the whole-brain network were defined based on the AAL template. Second, probabilistic tractography was used to estimate the connections between nodes in the whole-brain network (i.e. edges). Finally, the data was averaged across participants.

*Group averaged FC matrices*

fMRI signals were detrended and demeaned before band-pass filteringin the 0.04–0.07 Hz range (Glerean et al., 2012). The frequency range of 0.04–0.07 Hz was chosen because when mapped to the gray matter this frequency band was shown to contain more reliable and functionally relevant information compared to other frequency bands, and to be less affected by noise(Biswal et al., 1995, Glerean et al., 2012, Achard et al.,2006, Buckner et al.,2009). Subsequently, the filtered time series were transformed to z-scores. For each sleep stage, 15 participants were selected based on the presence of uninterrupted epochs of that sleep stage lasting more than 200 samples. Afterwards, the FC matrix was defined as the matrix of Pearson correlations between the fMRI signals of all pairs of regions of interest (ROIs) in the AAL template. To avoid confounds related to the length of the time series, the correlations were computed using only the first 200 volumes of each sleep stage. Fixed-effect analysis was used to obtain group-level FC matrices, meaning that the Fisher's R-to-z transform ($z = \tanh(R)$) was applied to the correlation values before averaging over participants within each of the sleep stages.

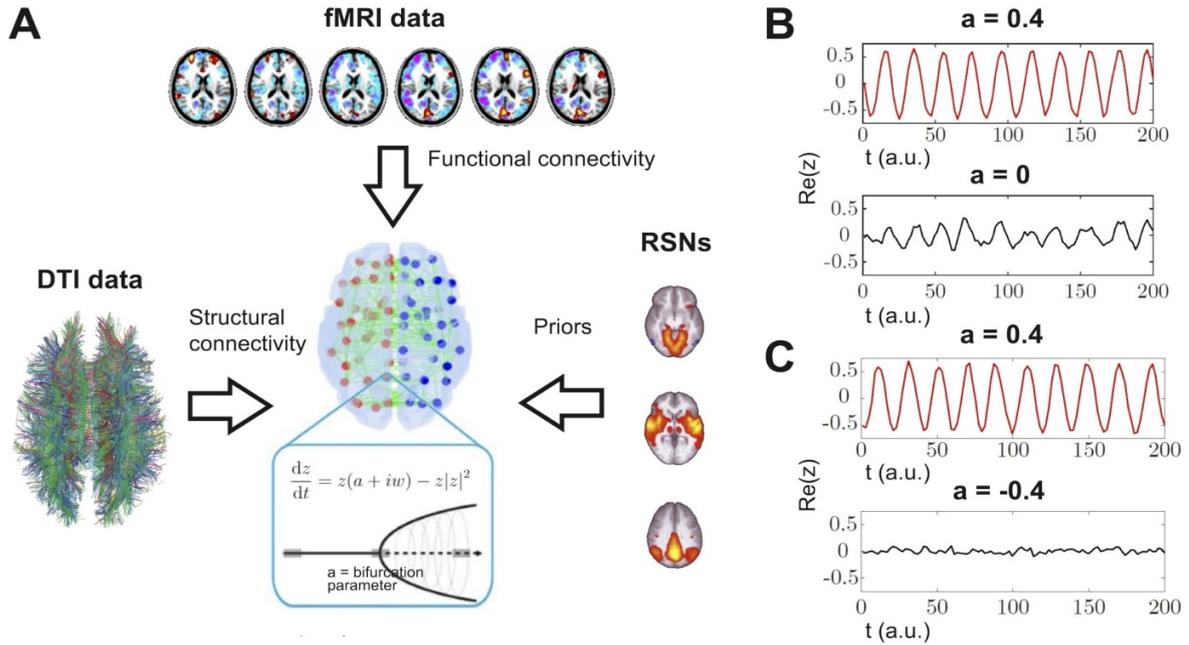

**Figure 1**: Procedure followed to construct the whole-brain model and a simplified example of the dynamics of two coupled oscillators. A) The model incorporates DTI data to define the SC between the non-linear oscillators, fMRI data to determine the intrinsic oscillation frequency of each node and the empirical FC that is fitted in the simulations, and RSNs as an anatomical prior to define the groups of nodes that contribute independently to the local bifurcation parameters. Dynamicsare given by the normal form of a supercritical Hopf bifurcation (the equations and bifurcation diagram are provided in the inset). B) Example of two coupled oscillators with different bifurcation parameters. The upper panel shows how oscillatory dynamics (a>0) can induce oscillations in a critical node (a=0) due to their coupling, while panel C shows that noisy dynamics at the fixed point (a>0) prevents the synchronization with the oscillating node (a<0). Re(z) stands for the simulated fMRI time series.

*Computational whole-brain model*

The general approach followed to construct the semi-empirical model is presented in the left panel of Fig. 1. Three different sources of empirical data were combined in a model where local dynamics are given by weakly interacting nonlinear oscillators. DTI data provided an estimate of SC between the oscillators, fMRI was used to estimate the intrinsic oscillation frequency of the local dynamics, and also provided empirical FC matrices that were used to fit the simulations, and RNSs determined a natural prior to group the nodes that contributed independently to the final local bifurcation parameters.

The implemented whole-brain model consists of a network of nonlinear oscillators coupled by SC. Each oscillator represents the dynamics at one of the 90 brain regions in the AAL template. The key neurobiological assumption is that dynamics of macroscopic neural masses can range from fully synchronous (i.e. activated state) to a stable asynchronous state governed by random fluctuations. A secondary assumption is that fMRI can capture the dynamics from both regimes (mediated by hemodynamic changes at a slower temporal scale, allowing to neglect the effect of signal transmission delays) with sufficient fidelity to be modeled by the equations.

Based on previous work (Jobst et al., 2017; Deco et al., 2017; Deco et al., 2018), the nonlinear oscillators were modeled by the normal form of a supercritical Hopf bifurcation. This type of bifurcation can change the qualitative nature of the solutions from a stable fixed point in phase space towards a limit cycle allowing the model to present self-sustained oscillations. While models of higher complexity could display analogous behavior, the normal form of a Hopf bifurcation was chosen for reasons of simplicity and generality, since it includes the minimal number of nonlinearities representing this range of dynamics.

Without coupling, the local dynamics of brain region j was modeled by the complex equation:

$$\frac{dz_j}{dt} = [a + i\omega_j]z_j - z_j|z_j|^2$$

(1)

In this equation z is a complex-valued variable ($z_j = x_j + iy_j$), and $\omega_j$ is the intrinsic oscillation frequency of node j. The intrinsic frequencies ranged from 0.04-0.07 Hz and were determined by the averaged peak frequency of the bandpass-filtered fMRI signals of each individual brain region.

The parameter a is known as the bifurcation parameter and controls the dynamical behavior of the system. For $a < 0$ the phase space presents a unique stable fixed point at $z_j = 0$, thus the system asymptotically decays towards this point. For $a > 0$ the stable fixed point changes its stability, giving rise to a limit cycle and to self-sustained oscillations with frequency $f_j = \omega_j/2\pi$ (Deco et al., 2017).

The coordinated dynamics of the resting state activity are modeled by introducing coupling determined by SC estimated from DTI. Nodes i and j are coupled by $C_{ij}$(the i,j entry of the SC matrix). To ensure oscillatory dynamics for a > 0, the SCmatrix was scaled to 0.2 (weak coupling condition) (Deco et al., 2017). In full form, the coupled differential equations of the model are the following:

$$\frac{dx_j}{dt} = \frac{dRe(z_j)}{dt} = [a - x_j^2 - y^2]x_j - \omega_j y_j + G \sum_i C_{ij}(x_i - x_j)$$
$$+ \beta\eta_j(t)$$

$$\frac{dy_j}{dt} = \frac{dIm(z_j)}{dt} = [a - x_j^2 - y^2]y_j + \omega_j x_j + G \sum_i C_{ij}(y_i - y_j)$$
$$+ \beta\eta_j(t)$$

(2)

The parameter G represents a global coupling factor that scales SC equally for all the nodes. These equations were integrated to simulate empirical fMRI signals using the Euler algorithm with a time step of 0.1. $\eta_j(t)$ represents additive Gaussian noise. When a is close to the bifurcation (a~0) the additive gaussian noise gives rise to complex dynamics as the system continuously switches between both sides of the bifurcation.

To illustrate the dynamical behavior of coupled nonlinear oscillators, we show the results obtained from two nodes in Fig. 1B and 1C.We present two representative cases: 1) one node (node 1) isin the dynamical regime of self-sustained oscillations (a = 0.4), and the other node (node 2) is close to the bifurcation (a~0); and 2) node 1 as before (a > 0) while node 2 isat the stable fixed point regime (a=-0.4). For both examples, the value of the couplingcoefficient $C_{12}$was set to 0.01. In the first case it is apparent that the coupling with synchronous dynamics at node 1 induces oscillations in node 2, while this does not happen when node 2 is at the fixed-point regime. This example illustrates how the bifurcation parameter, which is interpreted as determining the level of regional activation, relates to the FC between the nodes, i.e. FC increases for coupled nodes when at least one of them is in the oscillatory regime and the other close to the bifurcation.

*Priors for regional variations in the bifurcation parameters*

We extended previous modeling efforts by introducing additional parameters accounting for regional variations in the dynamical regime of the nodes. Introducing an independent bifurcation parameter for each individual node would result in a costly optimization procedure, as well as prone to overfitting the data. Thus, we considered priors to reduce the dimensionality by grouping the 90 AAL nodes into (potentially overlapping) sets. We consider n of such groups, $g_1, \ldots, g_n$, each contributing an independent amplitude $\Delta a_1, \ldots, \Delta a_n$ to the final bifurcation parameter of the node, which is computed as the linear combination:

$$a_i = \sum_{j=1}^{n} \Delta a_j 1_{g_j}(i)$$

(3)

where $1_{g_j}$ is the indicator function of set $g_j$ (i.e. 1 if the node belongs to $g_j$ and 0 otherwise).

By introducing priors, we constrain how different groups of nodes can contribute independently to the final bifurcation parameters, while allowing for regional variation depending on the precise definition of the sets $g_j$. In the Results section we explore four different priors: the heuristic prior (based on an intuitive grouping of the nodes based on inspection of the average empirical FC matrix during wakefulness), the RSN prior (grouping the nodes by RSN membership), the random prior (randomly assigning the nodes to the groups) and the homogeneous prior (only one group containing all the nodes).

*Goodness of fit: structure similarity index (SSIM)*

Once the coupling scaling factor G and the amplitudes $\Delta a_i$ of the n groups of nodes are determined, our model simulates time series that can be used to compute FC matrices to be compared with the empirical data. Our goal was to fit the model to the data recorded during different sleep stages by inferring the optimal amplitudes $\Delta a_j$. Different metrics can be used to determine the goodness of fit (GoF), such as the euclidean or correlation distance between the FC matrices, or the mean and standard deviation of the Kuramoto order parameter computed from the Hilbert transform of the time series (Jobst et al., 2017). We opted touse a metric that balances sensitivity to absolute (e.g. euclidean distance) and relative (e.g. correlation distance) differences between the FC matrices, termed structure similarity index (SSIM). This metric is based on three observables computed from a 2D array of values: the luminance, the contrast and the structure. The final distance between two arrays x and y is obtained from the product of the three terms $l(x, y)$, $c(x, y)$ and $s(x, y)$ related to each of the observables:

$$\text{SSIM}(x, y) = l(x, y)^\alpha \cdot c(x, y)^\beta \cdot s(x, y)^\gamma$$

$$l(x, y) = \frac{2\mu_x\mu_y + C_1}{\mu_x^2 + \mu_y^2 + C_1}$$

$$c(x, y) = \frac{2\sigma_x\sigma_y + C_2}{\sigma_x^2 + \sigma_y^2 + C_2}$$

$$s(x, y) = \frac{\sigma_{xy} + C_3}{\sigma_x\sigma_y + C_3}$$

(4)

The values of the exponent of each term are commonly set to $\alpha = \beta = \gamma = 1$, $C_1 = (0.01\,L)^2$, $C_2 = (0.03\,L)^2$ and $C_3 = C_2/2$, with L depending on the dynamic range of the matrix. For correlation matrices between -1 and 1, L is set to 1 (Wang et al., 2004). The SSIM ranges between 0 (lowest similarity) and 1 (highest similarity). The variables $\mu_x, \mu_y, \sigma_x, \sigma_y$ and $\sigma_{xy}$ are the local means, standard deviations, and cross-covariance for images *x, y* respectively. We define GoF = 1 – SSIM.

*Scaling of the structural coupling*

The parameter G was selected by exhaustively exploring the model using a homogeneous bifurcation parameter (i.e. the same bifurcation parameter for all nodes). The GoF between empirical and simulated FC for wakefulness was computed over a 100×100 grid in parameter space, with the bifurcation parameter a in the [-0.2,0.2] interval and G in the [0,3] interval. Before computing the simulated FC, the time series from the model were resampled to one sample per 2 s and bandpass filtered in the 0.04 - 0.07 Hz range. After averaging 50 independent runs we found the absolute minimum of GoF in $a = 0$ and $G = 0.5$, with a mean GoF of 0.7 (i.e. 30% similarity according to the SSIM). These results were used as initial conditions in the following model that incorporated regional variation in the bifurcation parameters, fixing G as 0.5 in further analyses.

*Genetic algorithm for parameter optimization*

After determining G, the n amplitudes $\Delta a_1, \ldots, \Delta a_n$ remain to be optimized. The exhaustive exploration of the parameter space is computationally prohibitive, thus we resorted to the application of a genetic algorithm to optimize the amplitudes. This stochastic optimization procedure is inspired in biological evolution, and is based on an algorithmic representation of natural selection consisting of letting the most adapted individuals prevail in the next generation, spreading the genes responsible for their better adaptation.

The algorithm starts with a generation of 10 sets of parameters ("individuals") chosen randomly close to zero, and generating a population of outputs with their corresponding GoF. A score proportional to the GoF is assigned to each individual. Afterwards, a group of individuals is chosen based on their score ("parents"), and the operations of crossover, mutation and elite selection are applied to them to create the next generation. These three mechanisms can be briefly described as follows: 1) elite selection occurs when an individual of a generation shows an extraordinarily high GoF in comparison with the other individuals, thus this solution is replicated without changes in the next generation; 2) the crossover operator consists of combining two selected parents to obtain a new individual that carries information from each parent to the next generation; 3) the mutation operator changes one selected parent to induce a random alteration in an individual of the next generation. In our implementation, 20% of the new generation was created by elite selection, 60% by crossover of the parents and 20% by mutation. A new population is thus generated ("offspring") that is used iteratively as the next generation until at least one of the following halting criteria is met: 1) 200 generations are reached (i.e. limit of iterations), 2) the best solution of the population remains constant for 50 generations, 3) the average along the last 50 generation of GoF relative variations is less than $1e^{-6}$.

The output of the genetic algorithm are the n optimal amplitudes $\Delta a_1, \ldots\ldots, \Delta a_n$ representing the independent contributions of the groups of nodes to the local bifurcation parameters, and the simulated FC with the highest GoF.

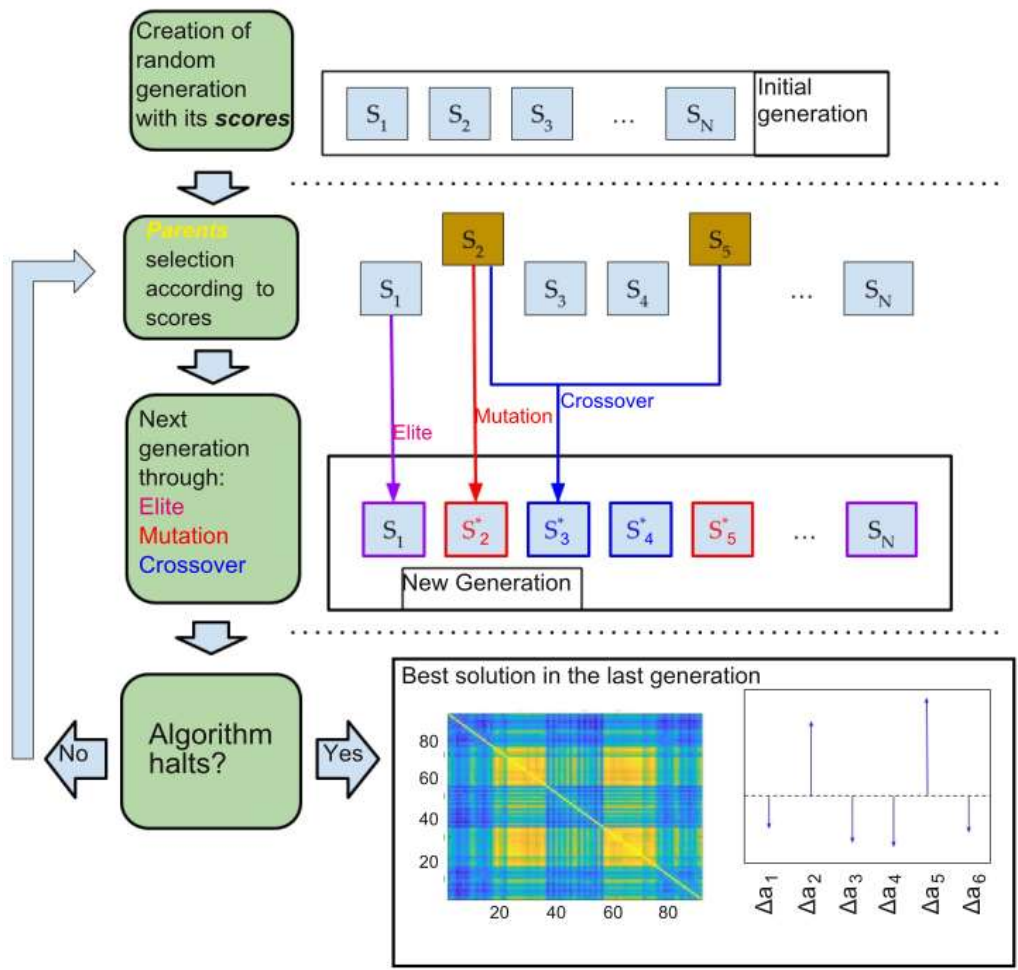

**Figure 2**: Schematic of the genetic algorithm implemented to optimize the group amplitudes. A population of 10 individuals (i.e. sets of parameters) with their corresponding scores (GoF of the empirical vs. simulated FC) is first generated, followed by a selection of parents based on their scores. A new generation of individuals is then generated by elite selection, crossover from the parents and mutation. This step is iteratively applied until at least one of the halting criteria is met. When finished, the algorithm outputs the optimal amplitudes together with the GoF and the simulated FC.

*Modeling transitions between brain states*

The optimization procedure was performed for the four brain states varying from awake to the deep sleep states using the RSNs priors. In each case six parameters were obtained, one perturbation per group, and the bifurcation parameters of each node were computed as is indicated in Eq. 3.

After obtaining the parameters leading to the optimal fit between empirical and simulated FC for wakefulness and all sleep stages, we modeled an external oscillatory stimulus and investigated whether it could induce a transition between deep sleep (N3) and wakefulness. The stimulus was represented as an external additive periodic forcing term incorporated to the equation of the j node, given by $F_j = F_{0_j} \cos(\omega_j t)$, where $F_{0_j}$ is the amplitude of the forcing and $\omega_j$ is the natural frequency of the node j. The effects of the forcing were investigated systematically for all 45 pairs of homotopic regions in the AAL atlas.

The forcing was initially applied for the optimal parameters to reproduce the deep sleep FC, with the forcing amplitude $F_{0_j}$ of node j and its homotopic pair being parametrically increased from 0 to 2 in steps of 0.05 (averaging 100 independent simulations for each node pair and $F_{0_j}$ value). For each value of $F_{0_j}$ the FC matrix was computed and its similarity with the wakefulness FC was determined as follows,

$$\Delta \text{GoF}_{norm} = \frac{\text{GoF}(FC_{sim_f}, FC_{emp_W}) - \text{GoF}(FC_{sim_W}, FC_{emp_W})}{\text{GoF}(FC_{sim_{N3}}, FC_{emp_W}) - \text{GoF}(FC_{sim_W}, FC_{emp_W})}$$

(5)

In this equation, $FC_{sim_f}$ is the FC matrix with forcing, $FC_{emp_W}$ the empirical FC matrix during wakefulness, $FC_{sim_W}$ the simulated FC matrix for wakefulness, and $FC_{si\ N3}$ the simulated matrix for N3 sleep (i.e. initial condition before the forcing is introduced). According to this normalization, as $\Delta\text{GoF}_{norm}$ approaches 0, the simulation starting from the optimal N3 bifurcation parameters plus the external forcing approaches the best empirical fit of the model to the wakefulness FC. Conversely, as $\Delta \text{GoF}_{\text{norm}}$ approaches 1 the forcingdoes not change the FC in the direction of the optimal FC for wakefulness.

**Results**

We first explored three different priors to determine the brain regions that contributed independently to the final local bifurcation parameters of the model. In the heuristic prior the nodes were grouped by visual inspection of the empirical FC matrix obtained during wakefulness (i.e. nodes presenting FC were included in the same group). In the RSN prior we assigned the nodes to groups based on membership to six different RSNs obtained from Beckmann et al., 2005 (Vis: primary visual cortex, ES: extrastriate cortex, Aud: regions associated with auditory processing, SM: sensorimotor regions, DM: default mode network, EC: executive control network). Note that a node can belong to more than one RSN; in that case, each of the overlapping RSNs contributes independently to the linear combination that yields the local bifurcation parameter. In the random prior we randomly assigned nodes to groups. These three priors result in models with six independent free parameters. Finally, in the homogenous prior all the nodes were assigned to the same group, therefore only one bifurcation parameter was used, as in Jobst et al., 2017. In the following we show and compare results obtained using different priors, and then apply the model using the best prior to obtain insights on possible mechanisms underlying the progressive transition from wakefulness to deep sleep.

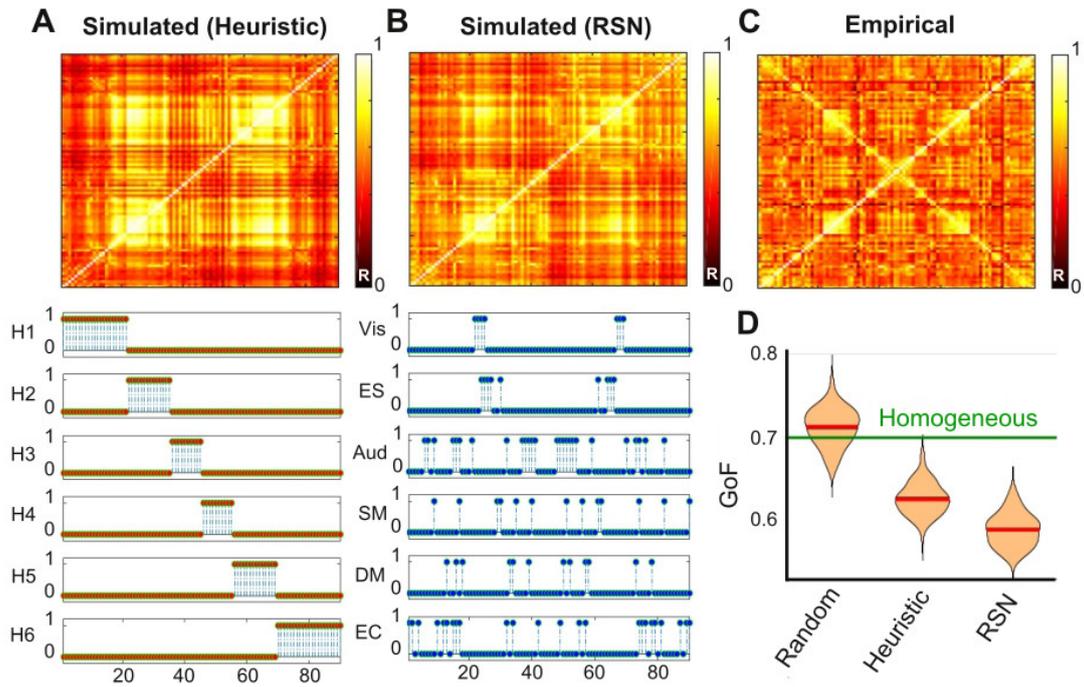

**Figure 3**: The best fit between empirical and simulated data was obtained using the RSN prior. FC matrices obtained from the whole-brain model fitted to fMRI FC using the heuristic prior (panel A) and the RSN prior (panel B). The bottom part of both panels shows the indicator function $1_{G_j}(i)$ signaling the group membership of node i. The empirical FC matrix is reproduced in panel C. As shown in panel D, the best GoF is obtained using the RSN prior, followed by the heuristic prior. Random assignment of the nodes to six groups does not outperform the homogeneous prior, in spite of having 5 more independent parameters.

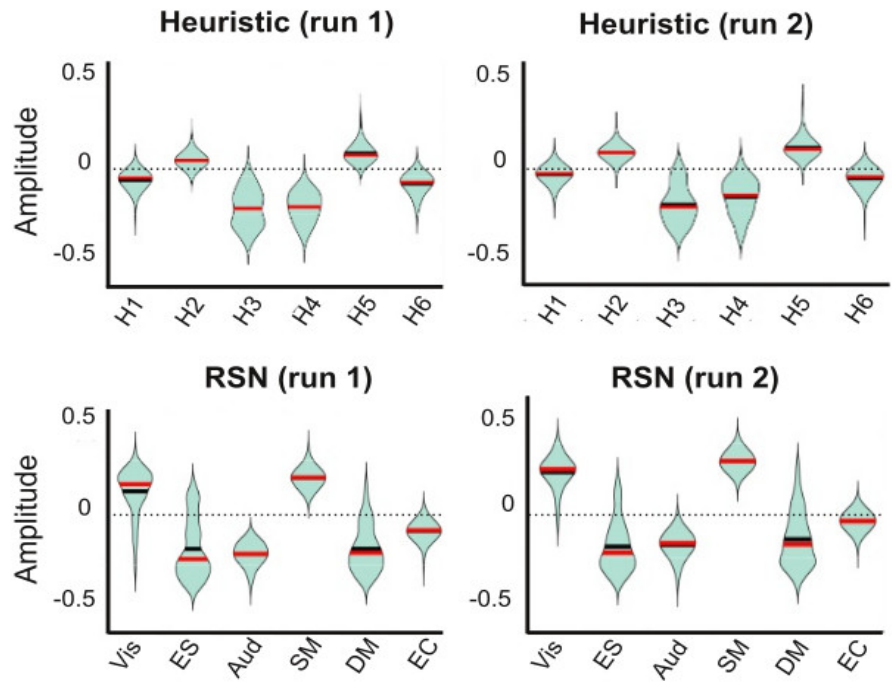

**Figure 4**: The amplitudes of each group of nodes yielding the optimal GoF between simulated and empirical FC for the heuristic prior (upper panel) and the RSN prior (bottom panel). Two sets of 100 independent runs were averaged and are shown to highlight the convergence of the model parameters. Black and red horizontal lines indicate the mean and the median of the distribution, respectively.

*Results of whole-brain modeling of wakefulness FC for the different priors*

Results obtained using the heuristic and RSN priors are compared in Fig. 3A and 3B, respectively. As seen upon visual inspection, in both cases the simulated FC approximated the empirical FC. The grouping of the nodes for both priors is shown in the bottom part of panels A and B. In particular, the sub-matrices corresponding to groups of nodes with high FC that appear on the diagonal and contradiagonal of both matrices (representing strong intra-hemispheric connections) are reproduced by both priors. The main divergence from empirical FC is manifested in the relatively small values on the contradiagonals, indicating that both models underestimated homotopic FC. This result was expected from the known underestimation of inter-hemispheric SC by DTI-based tractography (Deco et al., 2014; Messé et al., 2014; Reveley et al., 2015).

It is noteworthy that in spite of the *ad-hoc* selection of groups in the heuristic prior, grouping the node contributions in terms of RSNs resulted in a best GoF, as shown in Fig. 3D. It is also important to note that allowing independent regional contributions to the bifurcation parameters improves the GoF beyond what would be expected from simply increasing the number of free parameters in the model. This is because, as shown in Fig. 3D, random assignment of the amplitudes to 6 groups yielded a similar GoF to the model using the same bifurcation parameter for all nodes (homogeneous prior).

The optimal parameters for wakefulness FC shown in Fig 4 for the heuristic (upper panel) and RSN (bottom panel) priors. Results are shown averaged for two sets of 100 independent simulations to highlight the convergence of the genetic algorithm. The parameters shown in Fig. 4 correspond to the amplitudes that the nodes within each group contribute to the linear combination in Eq. 3, yielding the local bifurcation parameter. Thus, positive values (Vis and SM networks) indicate a contribution towards noise-dominated dynamics at the fixed point, while negative values (ES, Aud, DM, and EC networks) indicate a contribution towards oscillatory dynamics at the limit cycle.

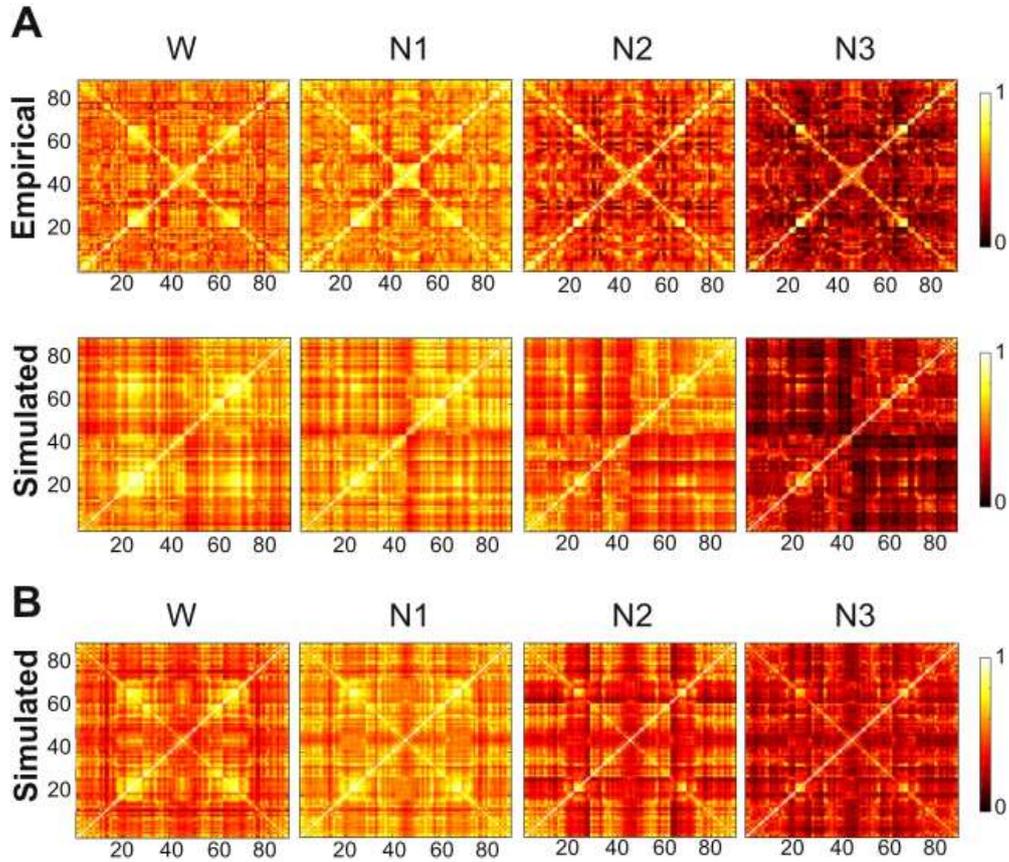

**Figure 5**: Comparison of empirical and simulated FC matrices for wakefulness and all sleep stages. A) Empirical and simulated FC, optimal fit using the RSN prior without changes to the SC. B) Same as panel A, but with a small *ad-hoc* increment in the homotopicSC

*Changes in regional dynamics from wakefulness to deep sleep*

Based on the results shown in Fig. 3B, we selected the RSNs as a canonical basis to regionally constrain the independent parameters in our model. Next, we fitted the whole-brain model to the empirical FC obtained during wakefulness, N1, N2 and N3 sleep. The estimated parameters correspond to the amplitudes that each node contributes to the linear combination yielding the final local bifurcation parameter.

The comparison of the optimal simulated FC matrices (average of 100 independent runs using the RSN prior) vs. the empirical FC is shown in Fig. 5. While for all stages the simulated intra-hemispheric FC resembled the empirical matrix, missing homotopic FC persisted during sleepand, for N3 sleep, the underestimation of inter-hemispheric FC extended to pairs of non-homotopic regions. Fig. 5B shows simulated FC obtained after adding a small *ad-hoc* positive value to homotopic SC, following previous work by Deco and colleagues (Deco et al., 2014) and Messé and colleagues (Messé et al. 2014). As expected from these previous works, this modification did not onlyimprove the simulated homotopic FC, but also the overall GoF by inducing a higher similarity between simulated and empirical non-homotopic inter-hemispheric FC.

Fig. 6 presents the parameters corresponding to the optimal simulated FC matrices presented in Fig. 5A. The upper panels display violin plots for the distribution of the amplitudes per RSN and sleep stage across 100 independent runs. The bottom panels present a comparison of these distributions for all pairs of stages in terms of Cohen's d ($d_{Cohen}$). Primary visual nodes (Vis network) contributed towards oscillatory dynamics during wakefulness, but this contribution progressively approached zero as the subjects transitioned towards N3 sleep. This appears reflected in the matrix containing the $d_{Cohen}$ values, since the effect size is in the "very large" range ($d_{Cohen} > 0.8$) for the comparison of N3 vs. all stages. Results were similar for sensorimotor regions (SM network). Conversely, default mode regions (DM network) approached zero during sleep from negative amplitude values, i.e. local dynamics unfolded around a stable fixed point during wakefulness, and approached the bifurcation progressively as the subjects transitioned towards N3 sleep. The remaining RSNsdid not present clear trends, as their amplitudes remained relatively stable around negative values (ES and Aud networks), or values very close to the bifurcation point (EC network).

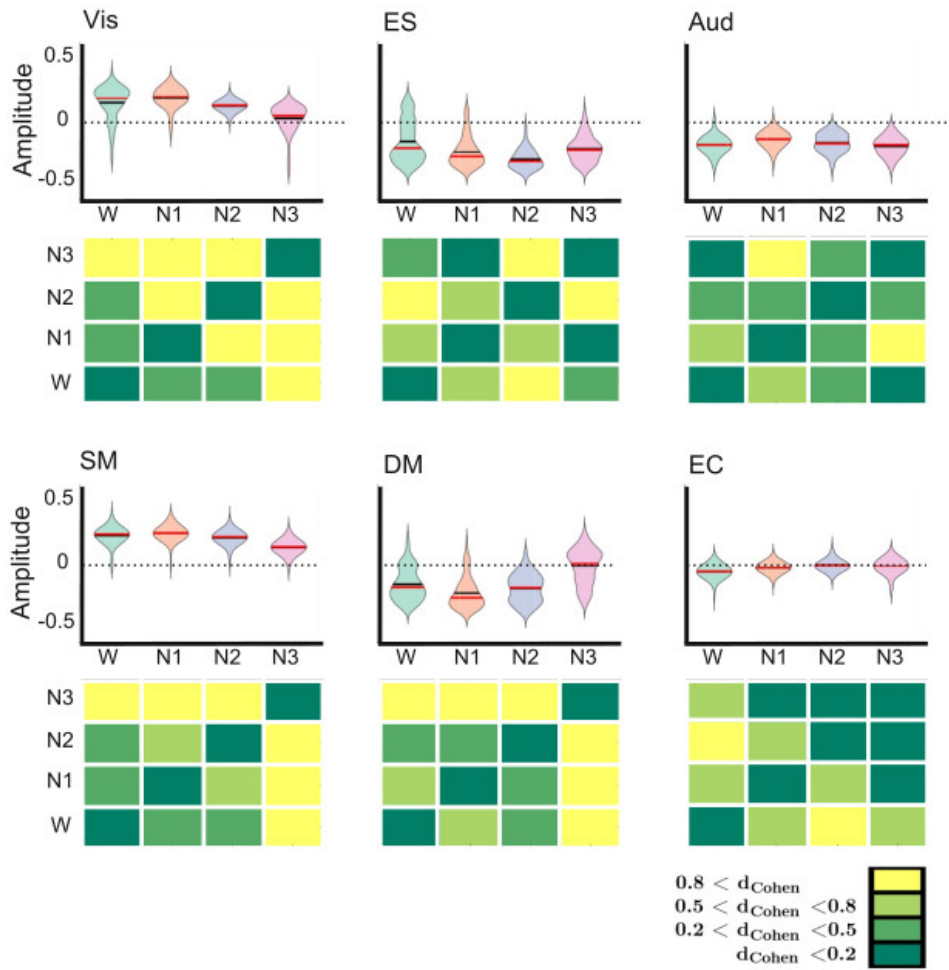

**Figure 6**: The amplitudes corresponding to the six RSNs, estimated from the optimal fit to the empirical FC data recorded during wakefulness (W), N1, N2 and N3 sleep. The bottom panels show Cohen's d ($d_{Cohen}$) for all pairwise comparisons. Primary visual (Vis) and sensorimotor (SM) nodes contributed towards oscillatory dynamics during wakefulness, but this contribution progressively approached zero as the subjects transitioned towards N3 sleep. The opposite result was observed for default mode (DM) nodes.

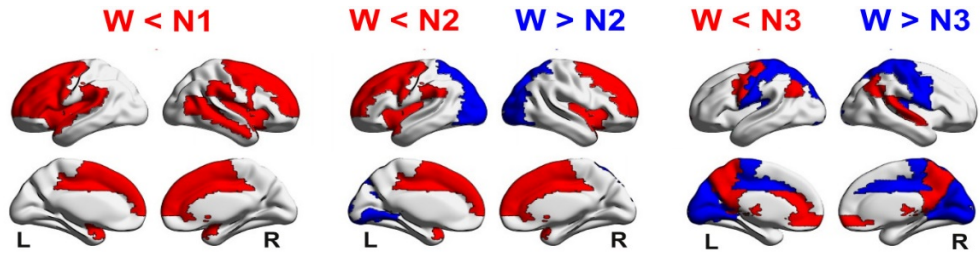

**Figure 7**: Changes in regional bifurcation parameters during sleep relative to wakefulness. A) Rendering of the regions associated with very large effect sizes ($d_{Cohen} > 0.8$) in the comparison of the bifurcation parameters corresponding to sleep (N1, N2 and N3) vs. wakefulness. Red and blue regions indicate $d_{Cohen} > 0.8$ for wakefulness < sleep and wakefulness > sleep, respectively. This implies that sleep transitions the dynamics towards a≈0, i.e. dynamics more susceptible to external perturbations.

For the nodes belonging to a single RSN, their amplitudes equal their bifurcation parameter; however, this does not need to be the case for multiple RSN memberships. The differences in the final bifurcation parameters (wakefulness vs. N1, N2 and N3 sleep) are shown rendered into brain anatomy in Fig. 7. Sleep reduced the bifurcation parameter in nodes belonging to sensory and motor networks, as well as to frontal regions in the case of N1 sleep. The opposite result was observed in parietal and frontal regions belonging to the DMN. This result was also present for N2 and N3 sleep vs. wakefulness, with the addition of frontoparietal and temporal nodes becomingshifted from fixed point towards oscillatory dynamics.

*Modeling the qualitative behavior of subcortical nodes during the wake-sleep transition*

The onset of sleep is known to bring about changes in the activity and FC of subcortical nuclei, especially those located within the thalamus and hypothalamus (Magnin et al., 2010; Picchioni et al., 2014; Tagliazucchi and Laufs, 2014). The thalamus consists of a multitude of nuclei of neurons densely connected by reciprocal pathways with the cerebral cortex, which have multiple functions, including acting as a relay station between sensory systems and cortex (Sherman and Guillery, 1996). In has been speculated that changes in thalamic activity during sleep onset could be related to the need of isolating the brain from external arousing stimuli (Magnin et al., 2010). Thalamic deactivation precedes cortical deactivation, leading us to expect changes in the bifurcation parameters of subcortical nuclei during N1 sleep.

We extended the RSN prior to include an additional set of nodes corresponding to subcortical structures within the AAL atlas. It is interesting to note that since these structures areincluded within several RSNs, their amplitudes and the final bifurcation parameters need not necessarily be equal. Thus, adding these nodes as an independent group allowed us to test whether we could reproduce known neurophysiological changes that occur at the onset of sleep, and to evaluate whether parameters that are 'hidden' within the model as part of the combination of variables determining the observable local dynamics can contain neurobiological meaningful information.

The results of this analysis are presented in Fig. 8. The left panel shows the amplitudes of the subcortical nodes as a function of sleep stage; it can be seen that the contribution of these nodes to the bifurcation parameter is consistently negative, with the largest negative value during N1 sleep. $d_{Cohen}$ values for the comparison between all sleep stages confirm that N1 sleep presents subcortical amplitudes different to those of wakefulness, N2 and N3 sleep. Since moving away from synchronized oscillatory dynamics is associated with decreased FC, this result is consistent with previous work showing that subcortical regions become deactivated and decoupled from the cortex during early sleep. The right panel shows that this result is less evident in the bifurcation parameters, with the effect sizes becoming smaller and reaching> 0.8 only for the comparison against N3 sleep. This suggests that information concerning the subcortical decoupling is more readily retrieved from hidden variables (amplitudes) than from the bifurcation parameters, which are directly related to the observables produced by the model.

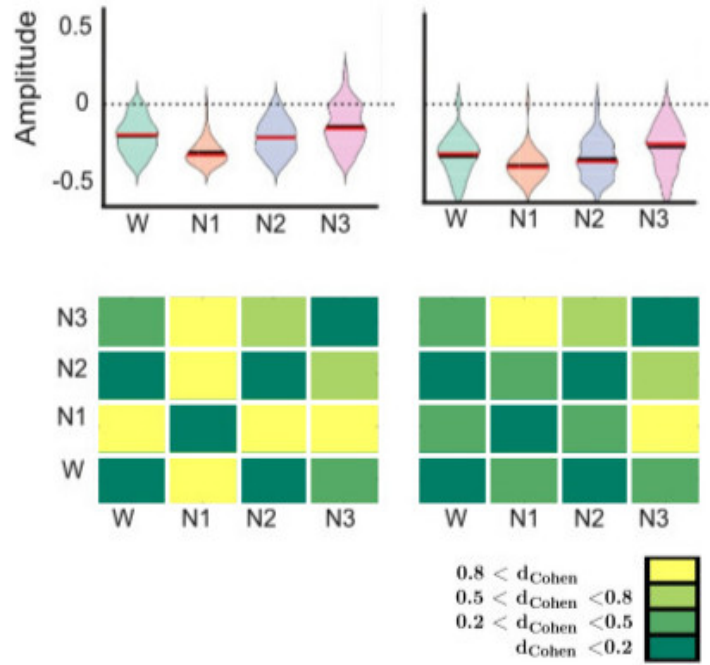

**Figure 8**: Changes in the amplitudes (left) and bifurcation parameters (right) of subcortical nodes from wakefulness to deep sleep

*Modelling externally induced transitions from deep sleep to wakefulness*

An important justification for the development of computational models of whole-brain activity is the *in silico* rehearsal of invasive or non-invasive brain interventions (e.g.transcranial alternating current stimulation [tACS]). Exploratory computational analyses could help identify optimal external perturbations to induce transitions between brain states.

We modified our model to simulate external perturbations with the aim of inducing an arousal from the deepest sleep stage (N3) to wakefulness. Our simulated stimulation protocol was based on an additive oscillatory forcing applied to pairs of homotopic nodes using the natural frequency of those nodes to maximize the effect of the perturbation. We exhaustively explored the values of the forcing amplitude $F_0$ from 0 to 2 in steps of 0.05, applied to all 45 pairs of homotopic nodes. The capacity of the stimulation to induce transitions between states was assessed using the metric $\Delta GoF_{norm}$ (see Methods section).

The qualitative behavior of $\Delta GoF_{norm}$ as a function of $F_0$ was heterogeneous and depended on the stimulated pair of nodes. We identified 10 pairs of nodes presenting $\Delta GoF_{norm}$ values below 0.5 for at least one value of $F_0$; in other words, by stimulating each of these 10 pairs of nodes, the simulated FC was closer to that of wakefulness than to deep sleep.

The behavior of these pairs of nodes against $F_0$ is shown in Fig. 9. Nodes presented three different qualitative behaviors; in the plot of $\Delta GoF_{norm}$ vs. $F_0$ these are indicated with different colors. In red, stimulation of the posterior cingulate cortex (PCC) monotonously decreased $\Delta GoF_{norm}$ up to 0.24. In green, the inferior occipital gyrus (IOG) presented an optimal $F_0$ value for which a minimum of $\Delta GoF_{norm}$ was reached (0.31) and afterwards remained approximately constant. Finally, the middle temporal gyrus (MTG, shown in blue) presented a clear global minimum of $\Delta GoF_{norm}$ (0.25) after which it increased as a function of $F_0$.

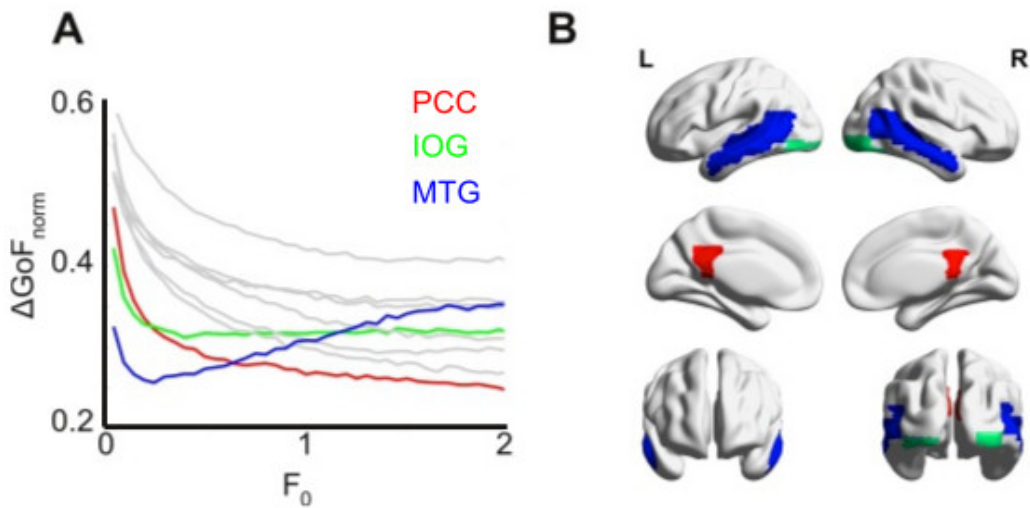

**Figure 9**: *In silico* stimulation of the model fitted to deep sleep using an additive oscillatory forcing term. A) $\Delta GoF_{norm}$ against the forcing amplitude $F_0$ for the 10 pairs of nodes leading to the lowest $\Delta GoF_{norm}$ values. B) Rendering of the three regions presenting the lowest $\Delta GoF_{norm}$; the color code indicates three different qualitative behaviors as $F_0$ is increased: $\Delta GoF_{norm}$ decreases as a function of $F_0$ (posterior cingulate cortex [PCC], shown in red), $\Delta GoF_{norm}$ achieves an optimal value and then increases as a function of $F_0$ (middle temporal gyrus [MTG], shown in blue), and $\Delta GoF_{norm}$ remains approximately constant as a function of $F_0$ (inferior occipital gyrus [IOG], shown in green).

**Discussion**

The usefulness of computational models in neuroscience -and more generally in the biological sciences- is related to a balance between model complexity and interpretability (Breakspear, 2007). A model should capture a series of conceptual factors and, by means of the freedom granted by *in silico* exploration, allow studying how the behavior of the system depends on these factors in isolation from the intractable complexity that characterizes living matter. Quoting John von Neumann, "*With 4 parameters I can fit an elephant, and with 5 I can make him wiggle his trunk*" (Dyson, 2004); meaning that the introduction of more parameters in a model should be adequately motivated lest it is rendered meaningless due to overfitting.

Previous studies applied supercritical Hopf bifurcations to model the local dynamics of whole-brain fMRI data (Deco et al., 2017a; Deco et al., 2017b; Jobst et al.,2017, Donnelly-Kehoe et al.,2019). Our work expanded the parameter space to account for regional variations in the level of activation, interpreted as the capacity of a node to engage in sustained large amplitude synchronous activity. The functional segregation of the human brain into systems that are differentially activated depending on task demands is known since the earliest days of neurology, and this knowledge was greatly advanced by the introduction of non-invasive neuroimaging tools (Frackowiak, 2017). While different brain states bring about global changes in brain activity, assuming global homogeneity is certainly an oversimplification. Thus, we approached the modeling of states in the wake-deep sleep progression by introducing new parameters representing the capacity of each chosengroup of nodes to drive the local dynamics towards or away from the Hopf bifurcation.

Our approach was agnostic with regard to the grouping of the nodes; however, we determined that the RSN prior overperformed an *ad-hoc* grouping based on inspection of the empirical FC matrix. The relevance of the anatomical priors is evident from the observation that, regardless of incorporating 5 new independent parameters, the random assignment of nodes to the groups led to a GoF similar to that obtained using the same parameter for all nodes (i.e. the homogenous prior). That the best fit was obtained when grouping the nodes by RSN membership does not only represent a useful development in terms of model building, but also yields insights on the natural way to describe the functional architecture of the brain. Thus, the incorporation of empirical anatomical priors in the modeling of other brain states could allow the evaluation of mechanistic hypothesis for the experimental observations. As an example, recent work by Deco and colleagues used maps of receptor density as priors to modulate parameters of a whole-brain model fitted to fMRI data acquired under the acute effects of a serotonergic psychedelic substance (Deco et al., 2018b).

The simplifications of our computational model are justified by assumptions whose validity should be discussed before engaging in the interpretation of the findings. Oscillatory dynamics are ubiquitous in electrophysiological data, and reconfigurations of the power spectrum of collective neural oscillations are a landmark feature of transitions between different brain states (Steriade et al., 1993; Buzsaki, 2006). While fMRI islimited with respect to the identification of synchronous dynamics due to its comparatively lowsampling rate, reliable reports of hemodynamic oscillations (supported by ultrafast fMRI sequences) exist (McAvoy et al., 2008; Baria et al., 2011). These oscillations are likely of neural origin, since multimodal EEG-fMRI studiesdemonstrated their positivecorrelation with infra-slow (0.01 – 0.1 Hz) oscillations in scalp potentials (Keinänen et al., 2018). More sophisticated models of mesoscopic brain activity, such as neural mass models, present several bifurcations including a supercritical Hopf from noisy to oscillatory dynamics (Grimbert and Faugeras, 2006; Coombes, 2010). By adopting the normal form of the bifurcation as a model, we made the further simplification that each node has a single dominant intrinsic oscillation frequency. Finally, our model neglects conduction delays since the temporal scale of fMRI oscillations is much slower than signal propagation times (Cabral et al., 2011).

The synergy between different empirical sources of data and our theoretical model for regional dynamics allows interpretability of the parameters beyond what could be inferred from fMRI data alone. For instance, while oscillatory dynamics in the fMRI signals could be assessed by computing the power spectrum (Baria et al., 2011), our model allows to disentangle the intrinsic oscillatory dynamics of each region from synchronization arising due to collective effects (i.e. coupling between nodes with different degree of proximity to the bifurcation; see the example provided in Fig. 1). Furthermore, our proposed model includes hidden variables, the amplitudes $\Delta a_i$, whose linear combination yields the bifurcation parameter, directly related to an empirical observable (the proximity to synchronous dynamics). However, as shown in Fig. 8 (left panel), these hidden variables represent neurobiologically relevant results, such as the known subcortical uncoupling and deactivation at sleep onset (Magnin et al., 2010) which can be obscured in terms of the resulting bifurcation parameters (Fig. 8, right panel). This result suggests that model fitting to individual fMRI and DTI data could yield additional parameters with valuable contributions towardsbrain state discrimination, with potential applications in the training of machine learning models for neurological disease diagnosis and prognosis.

The regional distribution of the estimated bifurcation parameters pictures the division of the cortex into two different dynamical regimes along the progression from wakefulness to deep sleep. Sensory areas approached the bifurcation fromoscillatory dynamics, while higher-level regions such as those in the DMN presented the opposite behavior. In terms of FC, these changes translated into decreased long-range correlationsbetween the simulated time series, consistent with multiple reports of regionally decreased FC during sleep (Horovitz et al., 2009; Sämann et al., 2011; Tagliazucchi and Laufs, 2014; Tagliazucchi and Van Someren, 2017). Recent work by Song and colleagues showed that the onset of sleep increases the power of BOLD oscillations throughout widespread cortical and subcortical regions (Song et al., 2019). An interesting observation arising from our work is that sleep gives rise to a state of diminished FC and -in several brain regions- decreased activation (quantified in terms of oscillation amplitude); however, by virtue of increased proximity to the Hopf bifurcation, these changes also endow sleep with higher synchronizability, i.e. the latent capacity to react upon external perturbations. In this sense, our model captures the difference between an on-line activated state with ongoing stable dynamics (wakefulness), and an off-line deactivated state with reduced and noisy intrinsic dynamics, but highly reactive to environmental perturbations (sleep).

A key factor influencing the interpretation of the model is the metric chosen to determine the GoF. Previous studies fitted metrics related to FC dynamics and metastability (Hansen et al., 2015; Deco et al., 2017; Orio et al., 2018), capturing the statistical distribution of FC temporal fluctuations. However, the optimal fit in this sense is not necessarily the optimal fit in the sense of reproducing the temporally averaged FC. A similar argument applies to GoF metrics based on the mean and variance of the Kuramoto order parameter. Since our aim was to investigate how regional dynamics related to inter-areal coordination during different brain states, the use of a GoF metric based on the static FC matrices emerged as a natural choice. However, future studies could optimize local parameters in terms of observables related to the level of metastability. Also, future models based on non-equilibrium dynamics (e.g. chaotic oscillators) (Li and Chen, 2004; Orio et al., 2018)could be explored as means to simultaneously describe static FC and its temporal fluctuations.

Our work provides an interesting example of how the effect of external oscillatory perturbations can be investigated via computational modeling, complementing previous work using other perturbation protocols (Deco et al., 2017; Deco et al., 2018, Saenger et al., 2017). The choice of periodic forcing aims to capture the effects of tACS, one of the most currently used and researched protocols for non-invasive electrical stimulation. The use of nodal natural oscillatory frequency (inferred from fMRI data) in the additive forcing term can be justified by reports of electrophysiological oscillations being entrained by in-phase tACS stimulation (Helfrich et al., 2014), even though this mechanism has been recently disputed (Lafon et al., 2017). Simulated periodic forcing at pairs of homotopic regions revealed non-trivial insights on the potential mechanisms underlying arousal from deep sleep. Regions within sensory systems (i.e. temporal and occipital lobes) appeared among those with the highest capacity to transition towards wakefulness; however, stimulation of the SC hub located at the posterior cingulate gyrus (Hagmann et al., 2018) resulted in FC with the highest resemblance to that of wakefulness. Furthermore, the relationship between forcing amplitude ($F_0$) and similarity with wakefulness ($\Delta GoF_{norm}$) was complex and region-specific. Sensory regions showed an optimal $\Delta GoF_{norm}$ at an intermediate value, presenting saturation or even diminishing $\Delta GoF_{norm}$ for increasing $F_0$, consistent with more biophysically realistic modeling by Ali and colleagues (Ali et al., 2013). In contrast, forcing at the posterior cingulate gyrus yielded a monotonously decreasing relationship between $\Delta GoF_{norm}$ and $F_0$. The rich connectivity of this node suggests that the effects of the forcing may depend on poly-synaptic connections reaching to other critical regions through one or more intermediate steps. Under this scenario, itbecomes critical that the development of non-invasive stimulation protocols to induce transitions between brain states is informed by computational models exhaustively exploring the effects of combinations at different brain regions. These models should incorporate information of the underlying individual SC, whose key role on the effect of external forcing is suggested by our results. In the same way we used a genetic algorithm combined with anatomical priors for dimensionality reduction to optimize the GoF, future studies could apply this method for the optimization of stimulation protocols to induce transitions, with potential applications in personalized medicine. An important caveat is that the source current density resulting from tACS stimulation bears a complex relationship with the scalp position of the electrodes, since intermediate tissues can distort the currents and a considerable fraction of the current leaks through conductive non-neural tissue (Kasinadhuni et al., 2017). Thus, the *in silico* rehearsal of stimulation protocols cannot prescind of personalized models of current propagation (Huang et al., 2017).

In conclusion, we implemented a computational model synthesizing different sources of empirical data to achieve a mechanistic description of intermediate complexity of the different brain states visited during the progression from wakefulness to deep sleep. This model led us to a number of insights narrowing the space of possible dynamical mechanisms allowing the stabilization and transition between self-organized brain states. We specifically addressed the conceptual validation of our model by contrasting its predictions with known neurobiological results. As a relatively simple and interpretable model whose flexibility and specificity emerges from the incorporation of empirical information, we expect our developmentwill find applications in the simulation of other brain states and -most crucially- in the rehearsal of protocols to induce transitions between them.

Deco, G., Cabral, J., Saenger, V. M., Boly, M., Tagliazucchi, E., Laufs, H., et al. (2018). Perturbation of whole-brain dynamics in silico reveals mechanistic differences between brain states. NeuroImage, 169, 46-56.

Deco, G., Cabral, J., Saenger, V. M., Boly, M., Tagliazucchi, E., Laufs, H., et al. (2018). Perturbation of whole-brain dynamics in silico reveals mechanistic differences between brain states. NeuroImage, 169, 46-56.

Deco, G., Cruzat, J., Cabral, J., Knudsen, G. M., Carhart-Harris, R. L., Whybrow, P. C., et al. (2018). Whole-brain multimodal neuroimaging model using serotonin receptor maps explains non-linear functional effects of LSD. Current biology, 28(19), 3065-3074.

Deco, G., Jirsa, V. K., & McIntosh, A. R. (2011). Emerging concepts for the dynamical organization of resting-state activity in the brain. Nature Reviews Neuroscience, 12(1), 43.

Deco, G., Jirsa, V. K., & McIntosh, A. R. (2013). Resting brains never rest: computational insights into potential cognitive architectures. Trends in neurosciences, 36(5), 268-274.

Deco, G., Kringelbach, M. L., Jirsa, V. K. & Ritter, P. The dynamics of resting fluctuations in the brain: metastability and its dynamical cortical core. Sci. Rep. 7, 3095, doi: 10.1038/s41598-017-03073-5 (2017).

Deco, G., McIntosh, A. R., Shen, K., Hutchison, R. M., Menon, R. S., Everling, S., et al. (2014). Identification of optimal structural connectivity using functional connectivity and neural modeling. Journal of Neuroscience, 34(23), 7910-7916.

Deco, G., Tagliazucchi, E., Laufs, H., Sanjuán, A., & Kringelbach, M. L. (2017). Novel intrinsic ignition method measuring local-global integration characterizes wakefulness and deep sleep. Eneuro, 4(5).

Demertzi, A., Antonopoulos, G., Heine, L., Voss, H. U., Crone, J. S., de Los Angeles, C., et al. (2015). Intrinsic functional connectivity differentiates minimally conscious from unresponsive patients. Brain, 138(9), 2619-2631.

Demertzi, A., Tagliazucchi, E., Dehaene, S., Deco, G., Barttfeld, P., Raimondo, F., et al. (2019). Human consciousness is supported by dynamic complex patterns of brain signal coordination. Science advances, 5(2), eaat7603.

Donnelly-Kehoe, P., Saenger, V. M., Lisofsky, N., Kühn, S., Kringelbach, M. L., Schwarzbach, J., ... & Deco, G. (2019). Reliable local dynamics in the brain across sessions are revealed by whole-brain modeling of resting state activity. Human brain mapping.